\documentclass[12pt,twoside]{article}   
    \usepackage[super,sort,comma]{natbib}

    \usepackage{fancyhdr}		
    
    \usepackage[section]{placeins}   %

    \usepackage{amsmath}
    \usepackage{amssymb}
    \usepackage{graphicx}
    \usepackage{subcaption}
    \usepackage{xcolor}
    \usepackage{authblk}
    \usepackage[nottoc]{tocbibind}
    \usepackage{siunitx}
    \usepackage{booktabs}
    
    \makeatletter \renewcommand\@biblabel[1]{$^{#1}$} \makeatother
    \usepackage[mathlines]{lineno}
    
    \usepackage{longtable}
    \usepackage{hyperref}
    \hypersetup{ colorlinks,
        citecolor=blue,
        filecolor=blue,
        linkcolor=blue,
        urlcolor=blue
    }
    
    \usepackage{xcolor}
    
    \definecolor{gray}{rgb}{0.6,0.6,0.6}
    \definecolor{red}{rgb}{0.85,0,0}
    \definecolor{green}{rgb}{0,0.85,0}
    \definecolor{blue}{rgb}{0,0,0.85}
    \definecolor{beige}{rgb}{0.92,0.87,0.78}
    \usepackage[all]{hypcap}    

\begin{document}


\title{Proton therapy range uncertainty reduction using vendor-agnostic tissue characterization on a virtual photon-counting CT head scan}


\author[1,2,3]{Stevan Vrbaski}
\author[1]{Goran Stanic}
\author[4]{Silvia Molinelli}
\author[3,5]{Mridul Bhattarai}
\author[3,5]{Ehsan Abadi}
\author[4]{Mario Ciocca}
\author[3,5]{Ehsan Samei}

\affil[1]{Vinaver Medical, 21000 Novi Sad, Serbia}
\affil[2]{Faculty of Medicine, University of Novi Sad, 21000 Novi Sad, Serbia}
\affil[3]{Carl E. Ravin Advanced Imaging Laboratories and Center for Virtual Imaging Trials, Duke University, Durham, NC 27705, USA}

\affil[4]{Centro Nazionale di Adroterapia Oncologica (CNAO), 27100 Pavia, Italy}
\affil[5]{Department of Radiology, Duke University, 2301 Erwin Rd, Durham, NC 27705, USA}

\maketitle

\section*{Abstract}

\noindent {\bf Background:} Photon-counting CT is the latest technology enabling imaging with reduced noise and inherent spectral separation, with the potential to directly calculate a more accurate tissue stopping power from spectral data. This potential benefit is difficult to quantify in practice and is currently evaluated mainly in phantoms with simplified geometries that only approximate real patient anatomy.\\ 
{\bf Purpose:}  In this work, we proposed virtual imaging simulators as an alternative approach to experimental validation of beam range uncertainty in complex patient geometry using a computational model of a human head and a CT system. In addition, we validate the accuracy of stopping power ratio (SPR) calculations on a model of a photon-counting CT scanner using a conventional stoichiometric calibration approach and a prototype software \textit{TissueXplorer}. \\
{\bf Methods:} A validated CT simulator (\textit{DukeSim}) was used to generate photon-counting CT projections of a computational head phantom, which were reconstructed with an open-source toolbox (\textit{ASTRA}). The dose of 2 Gy was delivered through protons in a single fraction to target two different cases of nasal and brain tumors with a single lateral beam angle. Ground-truth treatment plan was made directly on the computational phantom using clinical treatment planning software (\textit{RayStation}). This plan was then recalculated on the corresponding CT images for which SPR values were estimated using both the conventional method and the prototype software \textit{TissueXplorer}. The resulting dose distributions were subsequently compared against the ground-truth plan to quantify dose differences arising from SPR estimation.\\
{\bf Results:} The mean percentage difference in estimating the stopping power ratio with \textit{TissueXplorer} in all head tissues inside the scanned volume was 0.28\%. Stopping power ratios obtained with this method showed smaller dose distribution differences from the ground truth plan than the conventional stoichiometric calibration method on the computational head phantom.\\
{\bf Conclusions:}  Virtual imaging offers an alternative approach to validation of the SPR prediction from CT imaging, as well as its effect on the dose distribution and thus downstream clinical outcomes. According to this simulation study, software solutions that utilize spectral information, such as \textit{TissueXplorer}, hold promise for more accurate prediction of the stopping power ratio than the conventional stoichiometric approach.\\

\section*{Keywords}
stopping power ratio; TissueXplorer SPR; photon-counting CT; proton therapy; radiotherapy treatment planning; virtual imaging trials; anthropomorphic phantoms

\section{Introduction}

The physical characterization of the patient using computed tomography (CT) imaging leads to combined range uncertainties on the order of $\sim$3–3.5\% of water-equivalent path length, with representative experimental measurements indicating values of $\sim$1.6\% for soft and up to $\sim$5\% in lung and bone tissues \cite{yang_comprehensive_2012, hunemohr_experimental_2013, sarkar_evaluation_2023}. The major contributor for conventional single-energy CT (SECT) scanners is the conversion of Hounsfield units (HU) to tissue stopping power ratio (SPR) via a stoichiometric calibration curve \cite{moyers2020physical}, with variability between centers introduced by scanner-specific calibration protocols \cite{peters_experimental_2021}. Other contributions come from CT imaging noise and artifacts, and estimation of mean excitation energy. Dual-energy CT (DECT) has been extensively studied as a main candidate to reduce range uncertainty. Wohlfahrt et al. provided a comprehensive review on technical solutions developed for translating spectral information to SPR \cite{wohlfahrt_status_2020}. DECT-based SPR estimation has shown agreement with ground truth within 2\% \cite{sarkar_evaluation_2023} and 0.7\% \cite{longarino_potential_2022} for animal and tissue equivalent inserts. Nasmark et al. concluded that range uncertainty due to CT imaging noise is small compared to other contributions, but it becomes more important as we move towards smaller treatment margins \cite{nasmark_influence_2023}. Photon-counting CT (PCCT), offering reduced noise and inherent spectral separation, is the latest evolution in this direction. Zimmerman et al. reported improved estimation of relative electron density on a clinical PCCT scanner versus a DECT scanner, but comparable SPR accuracy \cite{zimmerman_assessment_2025}. Similar performance between SECT, DECT, and PCCT in estimating SPR values was observed in another independent phantom study \cite{huijskens2025photon}.

To account for uncertainty in physical characterization, most centers quantify residual SPR error from phantom measurements, propagate it as range uncertainty along each beam path, and incorporate it as a generic range margin during robust optimization.  In a neuro-oncology patient cohort, reducing the robust range uncertainty from 3\% to 2\% enabled by DECT-based SPR maps lead to clinically acceptable plans with organ-at-risk (OAR) dose reductions in 89\% of patients \cite{taasti_clinical_2023}. It has been shown that the difference in mean proton range at 90\% distal dose point was 0.1 mm (0.07\%) for DECT-based versus 2.2 mm (1.5\%) for SECT-based planning \cite{sarkar_evaluation_2023}. While measurement of average proton range provides a quantitative comparison, this approach gives information about the lump effect of overall absolute beam uncertainty and cannot distinguish between uncertainties in specific tissues or provide the resulting three-dimensional dose mismatch. Reliable in-vivo range metrology, such as ion radiography \cite{fogazzi_proton_2024}, is under development, but is not yet in clinical use.

The work presented in this paper has therefore two objectives. Firstly, we propose virtual imaging simulators as an alternative approach to experimental validation of beam range uncertainty. The motivation behind this objective is two-fold: i) virtual imaging enable absolute comparison of 3D dose distribution against (user-defined) ground truth in complex patient anatomy \cite{segars20104d,abadi_dukesim_2019}, and ii) support simulation of the newest CT technologies, including vendor-neutral virtual PCCT scanners based on various photon-counting detector designs \cite{sharma2024framework}. Secondly, we use this framework to validate a vendor-agnostic software prototype (\textit{TissueXplorer}, Vinaver Medical) developed to provide highly accurate SPR maps from PCCT scans in clinical conditions and compare the results with the ground truth and the stoichiometric SECT approach.

\section{Methods}

\subsection{Virtual imaging framework}

\textit{DukeSim} is a virtual imaging CT simulator that takes a voxelized computational phantom as input and reproduces the geometry and physics of a clinical system. The scanner geometry includes a modification of the X-ray spectrum and bowtie filtering, focal‐spot motion, tube-current modulation, and many other relevant parameters. The simulator supports both energy-integrating and photon-counting detector models and has been validated against commercial scanners \cite{abadi_dukesim_2019,Abadi2019ScannerSpecific,Jadick2021TCM,Shankar2022TaskBased,McCabe2025InSilico}.

For this work, a single-source, helical CT acquisition was simulated. The scanner geometry and spectrum corresponded to a clinically utilized 120 kVp configuration defined by a generic geometry file. The scan used 2016 projections per rotation, 0.5 s rotation time, a pitch of 0.8, and 5 rotations to cover the volume. Tube current was 250 mA with tube-current modulation enabled. X-ray scattering was modeled via the MCGPU module set to 5 $\times 10^8$ photon histories at 36 equidistant scatter angles/projections per rotation. Computations ran on 4 GPU nodes, and each simulation took approximately 2.5 hours. Detector was a generic photon-counting detector with a CdTe sensor with a pixel size of 0.5 $\times$ 0.6 mm and a thickness of 3 mm. Projections generated through this setup were separated into the low and high energy bins with energy thresholds at 20 keV and 65 keV by applying the corresponding detector. Spectral images were reconstructed in the DICOM format from corresponding projections with the open-source ASTRA toolbox using the GPU-enabled filter-back projection algorithm at matrix size 512×512, and an analytical (ramp) kernel. Image-domain virtual monochromatic images were generated using algorithms described elsewhere \cite{yu2012dual}.

This simulation set up was used to image two computational models: an anthropomorphic XCAT head phantom \cite{segars20104d} and a virtual model of the CIRS electron density and tissue-equivalent phantom (CIRS Inc.). The XCAT head phantom was generated using the tissue compositions and densities given in Table \ref{tab:composition}. Both phantoms had isotropic voxels of 0.5 mm$^3$. 

\begin{table}[ht!]
\centering
\small
\resizebox{\textwidth}{!}{%
\begin{tabular}{lccccccccccccc}
\hline
Tissue / Atomic number & 1 & 6 & 7 & 8 & 11 & 12 & 15 & 16 & 17 & 19 & 20 & 26 & Density \\
\hline
Air            & 0.000 & 0.000 & 0.755 & 0.232 & 0.000 & 0.000 & 0.000 & 0.000 & 0.013 & 0.000 & 0.000 & 0.000 & 0.001 \\
Adipose        & 0.120 & 0.640 & 0.008 & 0.229 & 0.000 & 0.000 & 0.002 & 0.000 & 0.000 & 0.000 & 0.000 & 0.000 & 0.916 \\
Laryngo-pharynx& 0.102 & 0.143 & 0.034 & 0.710 & 0.001 & 0.000 & 0.002 & 0.003 & 0.001 & 0.004 & 0.000 & 0.000 & 1.059 \\
Cortical bone  & 0.034 & 0.155 & 0.042 & 0.435 & 0.001 & 0.002 & 0.103 & 0.003 & 0.000 & 0.000 & 0.225 & 0.000 & 1.920 \\
Spine          & 0.063 & 0.261 & 0.039 & 0.436 & 0.001 & 0.001 & 0.061 & 0.003 & 0.001 & 0.001 & 0.133 & 0.000 & 1.374 \\
Spinal cord    & 0.111 & 0.000 & 0.000 & 0.880 & 0.005 & 0.000 & 0.000 & 0.000 & 0.004 & 0.000 & 0.000 & 0.000 & 1.010 \\
Bone marrow    & 0.082 & 0.393 & 0.029 & 0.373 & 0.000 & 0.000 & 0.039 & 0.000 & 0.000 & 0.000 & 0.083 & 0.000 & 1.194 \\
Blood          & 0.102 & 0.110 & 0.033 & 0.745 & 0.001 & 0.000 & 0.001 & 0.002 & 0.003 & 0.002 & 0.000 & 0.001 & 1.060 \\
Cartilage      & 0.096 & 0.099 & 0.022 & 0.744 & 0.005 & 0.000 & 0.022 & 0.009 & 0.003 & 0.000 & 0.000 & 0.000 & 1.112 \\
Gland          & 0.106 & 0.284 & 0.026 & 0.578 & 0.000 & 0.000 & 0.001 & 0.002 & 0.002 & 0.001 & 0.000 & 0.000 & 1.030 \\
Eye            & 0.096 & 0.196 & 0.057 & 0.649 & 0.000 & 0.000 & 0.001 & 0.000 & 0.000 & 0.000 & 0.000 & 0.000 & 1.070 \\
Brain          & 0.107 & 0.095 & 0.018 & 0.767 & 0.002 & 0.000 & 0.003 & 0.002 & 0.003 & 0.003 & 0.000 & 0.000 & 1.040 \\
\hline
\end{tabular}%
}
\caption{Elemental composition (mass fractions) and density for the XCAT head phantom.}
\label{tab:composition}
\end{table}

The virtual CIRS phantom material file has been replicated using vendor-provided material composition of inserts for air, solid water, cortical bone, LN450 lung-equivalent, brain-equivalent, breast (50/50 adipose–glandular), liver-equivalent, calcium carbonate (30\%), inner bone, calcium carbonate (50\%), and general adipose tissue.

\subsection{Ground truth SPR map}\label{sec:gt_prep}

The existence of ground truth is one of the main advantages of the virtual imaging framework. The ground truth for the XCAT phantom is established by associating each voxel in the phantom with chemical composition and density (\emph{e.g.,} brain XCAT in Table \ref{tab:composition}). The information about the composition and density is then used by the ray tracing and Monte Carlo modules of \textit{DukeSim} CT simulator to model X-ray interaction with each phantom voxel and compute the total (primary and scatter) detected signal in each detector pixel. The same information about composition and density for each material can be used to theoretically compute SPR values in each voxel. This information was considered to be the ground truth, to the extent that the Bethe-Bloch formula can be used to calculate the SPR. The Bethe-Bloch formula to calculate SPR is given in the equation \ref{eq:spr},

\begin{equation}
\mathrm{SPR}
=\frac{\rho \sum_i \frac{W_i\,Z_i}{A_i}}{\rho_{\mathrm{w}} \sum_j \frac{W_j\,Z_j}{A_j}}
\cdot
\frac{\ln\!\Big(\dfrac{2 m_e c^{2}\beta^{2}}{(1-\beta^{2})\, I_{\mathrm{comp}}}\Big)-\beta^{2}}
{\ln\!\Big(\dfrac{2 m_e c^{2}\beta^{2}}{(1-\beta^{2})\, I_{\mathrm{w}}}\Big)-\beta^{2}},
\label{eq:spr}
\end{equation}

where $\beta$ is the particle's velocity relative to the speed of light, 
$\gamma$ is the Lorentz factor:

\[
\beta \;=\; \sqrt{1-\frac{1}{\gamma^2}}, \qquad
\gamma \;=\; 1+\frac{E_{\mathrm{MeV}}\times 10^{6}\,e}{m_p c^{2}}
\]
$E_{\mathrm{MeV}}$ is the kinetic energy in MeV, $e$ is the elementary charge, $m_p$ is the proton mass, $c$ is the speed of light and $I$ is the mean ionization energy in joules:
\[
I_{\mathrm{comp}}
=\exp\!\left(
\frac{\sum_i \frac{W_i Z_i}{A_i}\,\ln I_i^{(\mathrm{eV})}}
     {\sum_i \frac{W_i Z_i}{A_i}}
\right)\,e,
\qquad
I_{\mathrm{w}}
=\exp\!\left(
\frac{\sum_j \frac{W_j Z_j}{A_j}\,\ln I_j^{(\mathrm{eV})}}
     {\sum_j \frac{W_j Z_j}{A_j}}
\right)\,e.
\]
Subscripts "water" and $j$ are associated with water, while "comp" and $i$ refer to any compound. $W_i$,  $W_j$ are the mass fraction of an element, $Z_i$, $Z_j$ are the atomic number of an element, $A_i$, $A_j$ are the atomic mass of an element (g/mol) in the compound and water, respectively. The $E_{\mathrm{MeV}}$ in this work was 100 MeV.

To compare ground-truth SPR maps with reconstructed DICOM CT images, the ground-truth phantom as a binary map was loaded, rotated, and flipped to match the DICOM patient orientation, and assigned an isotropic spacing of 0.5\, mm. The volumes were aligned by matching their physical centers and correcting the residual superior–inferior offset computed from center-of-image coordinates, after which the ground-truth volume origin was updated accordingly. Using the DICOM image as a reference, the ground-truth map was resampled (SimpleITK, ResampleImageFilter, Python v3.10) to the DICOM grid, preserving its spacing and direction, and then interpolated using nearest-neighbor interpolation to maintain discrete material labels. Finally, labels were converted to quantitative SPR by indexing a material lookup table \ref{tab:composition} and using the equation \ref{eq:spr}.

\subsection{TissueXplorer SPR}\label{sec:tissueXplorer}

\textit{TissueXplorer} is a prototype system for tissue characterization from spectral CT scans. Spectral CT systems from different vendors typically encode spectral information in multiple forms, including tissue mass or electron density, effective atomic number, low- and high-energy mixture images, or virtual monochromatic images. Virtual monochromatic images are the standard output of the first clinical PCCT scanner (NAEOTOM Alpha, Siemens Healthneers) \cite{rajendran2022first}, and can be obtained from other DECT scanners from major vendors. \textit{TissueXplorer} performs dictionary-based tissue classification from a virtual monochromatic data set and estimates the full material description for each pixel in the reconstructed slices. Information about the material density in each voxel is determined directly from the spectral CT scan. The dictionary contains attenuation properties of a large number of known tissues and their mixtures, including mean ionization energies and corresponding SPRs determined from these material properties. Dictionary materials were determined by interpolation from established reports on human tissue and metal core materials \cite{ICRU44_1989,ICRP110_2009,NIST_XCOM_2004}. The output of \textit{TissueXplorer} is either voxel-based tissue chemical composition and density or voxel-based SPR values.

\subsection{Treatment plan setup}

Treatment plans were optimized with \textit{RayStation} (Raysearch Laboratories), the commercial treatment planning software in clinical use at the Centro Nazionale di Adroterapia Oncologica (CNAO, Pavia, Italy) for two representative cases of nasal and brain tumors, chosen as challenging sites due to their proximity to critical structures such as the eyes, brain stem, optic nerves, and chiasm. The tumor material composition has not been altered from that of the healthy tissue in the delineated regions. Dose–volume constraints for the OARs were applied according to institutional standards, and robust optimization with pencil beam scanning was performed to account for range and setup uncertainties. A prescription dose of 2 Gy (RBE) to clinical target volume (CTV), corresponding to a typical applied treatment fraction, was used for the target with a single lateral beam angle.

In total, two plans (nasal and brain tumor) were created for three different scenarios. First, the ground-truth plan was created using the XCAT anthropomorphic phantom with ground-truth SPR values. Then, the single energy PCCT (SE-PCCT) plan and \textit{TissueXplorer} PCCT plan were recalculated from the original ground-truth plan, maintaining the treatment plan setup. The recalculated plans were compared to the ground truth by producing a percent difference map for each voxel using the formula:

\begin{equation}
    \% \, \textit{diff} = \frac{\text{Ground-truth  - PCCT}}{\text{PCCT}}\times 100
\label{eq:diff}
\end{equation}

Such comparison was possible because the ground-truth map was precisely scaled to the CT reconstructed volumes as described in Section \ref{sec:gt_prep}. Both the SE-PCCT and \textit{TissueXplorer} PCCT plans originate from PCCT scans, but the input for the SE-PCCT plan was CT reconstructions at a threshold of 20 keV, while the input for the \textit{TissueXplorer} PCCT plan was the virtual monochromatic CT reconstructions. The CIRS phantom has been used to create the stoichiometric calibration curve between the obtained Hounsfield units (HU), \textit{DukeSim}-generated reconstructions (120 kVp tube potential and 20 keV threshold reconstructions) and mass density (given by the manufacturer) for the SE-PCCT plan. The calibration curve was obtained by measuring mean values in regions of interest inside CIRS inserts on an artifact-free central slice. The obtained calibration curve has been commissioned following the standard clinical protocol of including a new SECT scanner in the treatment planning software. The PCCT plan was obtained through \textit{TissueXplorer} and did not require the stoichiometric calibration curve. The \textit{TissueXplorer} has been set to output SPR maps directly, and the commissioning into the treatment planning software was performed by constructing a line that scaled the imported values to absolute SPR values \cite{janson_treatment_2024}.

\section{Results}

\begin{figure}[ht!]
  \centering
  \subfloat[Virtual CIRS phantom \label{fig:gammex}]{
    \includegraphics[width=0.45\linewidth]{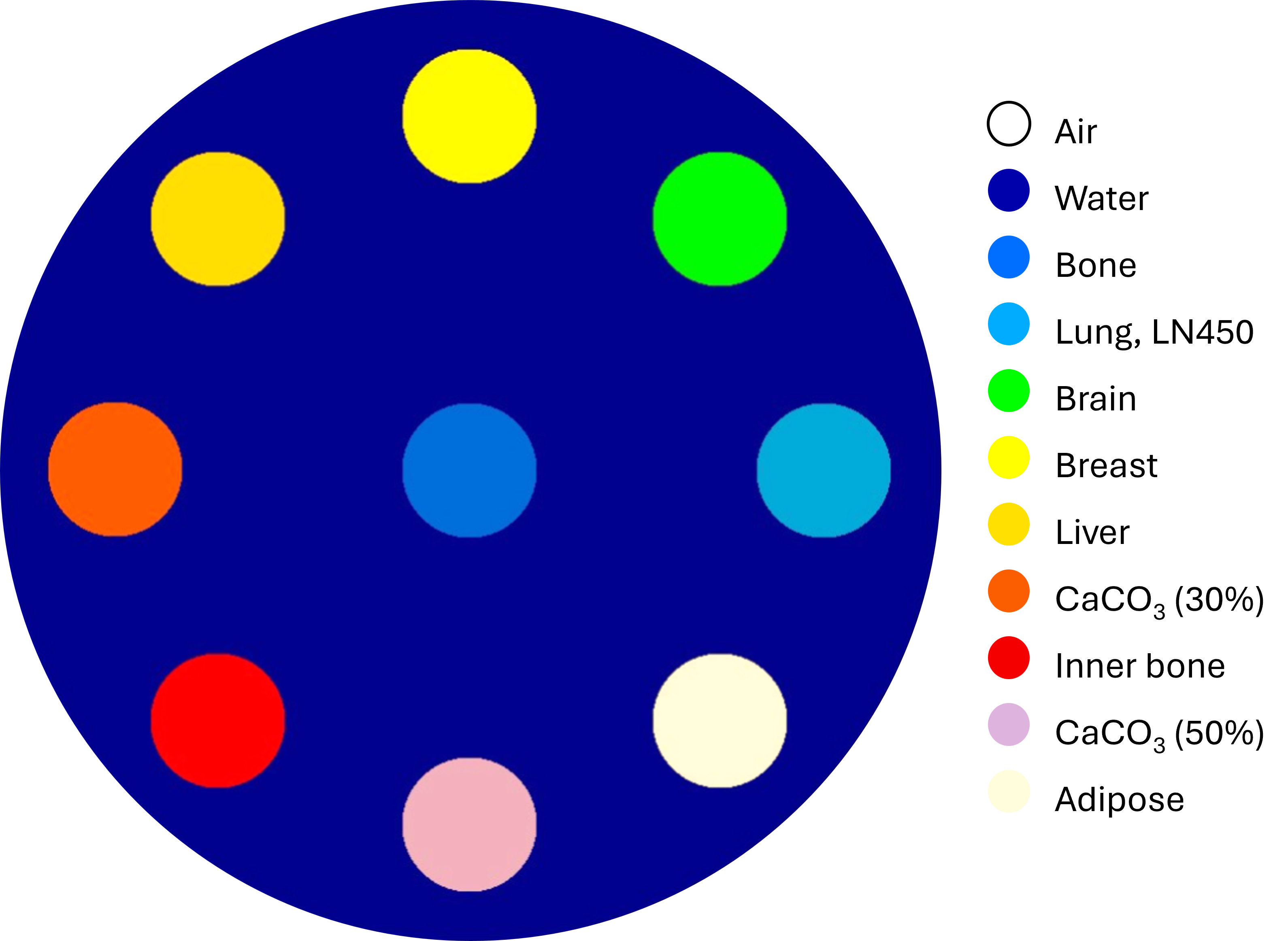}
  }\hfill
  \subfloat[XCAT head phantom\label{fig:xcat}]{
    \includegraphics[width=0.45\linewidth]{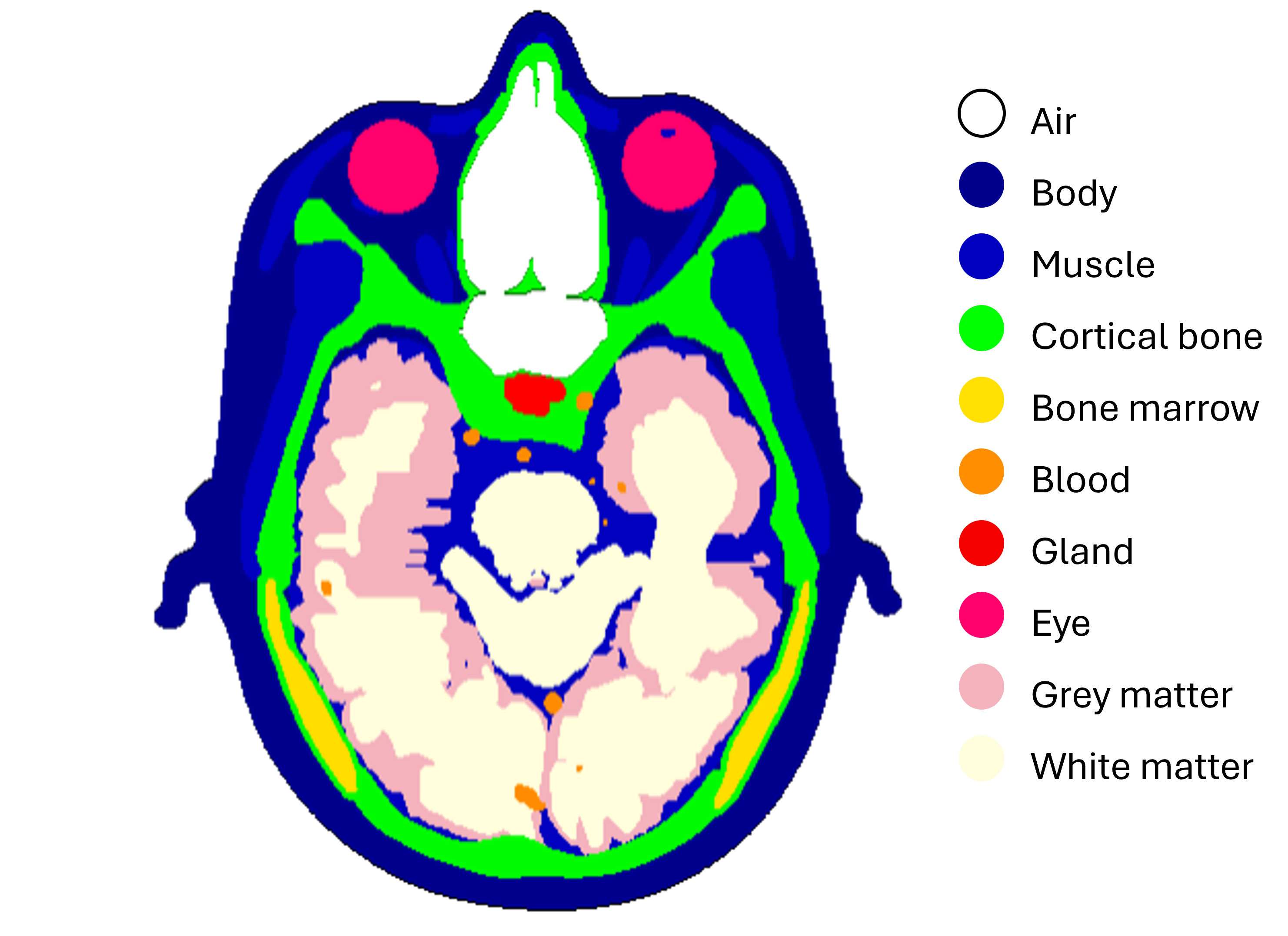}
  }
  \caption{Computational (virtual) binary phantoms used for (a) generating a calibration curve and (b) treatment planning with placeholders (marked with different colors) for materials of different composition.}
  \label{fig:phantoms}
\end{figure}

The CIRS and XCAT head computational phantoms used in this study are shown in figure \ref{fig:phantoms}. Phantoms are color-coded so that each color corresponds to a different tissue, as shown in the legend. Material compositions and densities are input to the \textit{DukeSim} simulator. In this work, we used vendor-provided compositions for the CIRS phantom and Table \ref{tab:composition} for the XCAT head phantom. 

\begin{figure}[ht!]
    \centering
    \includegraphics[width=0.8\linewidth]{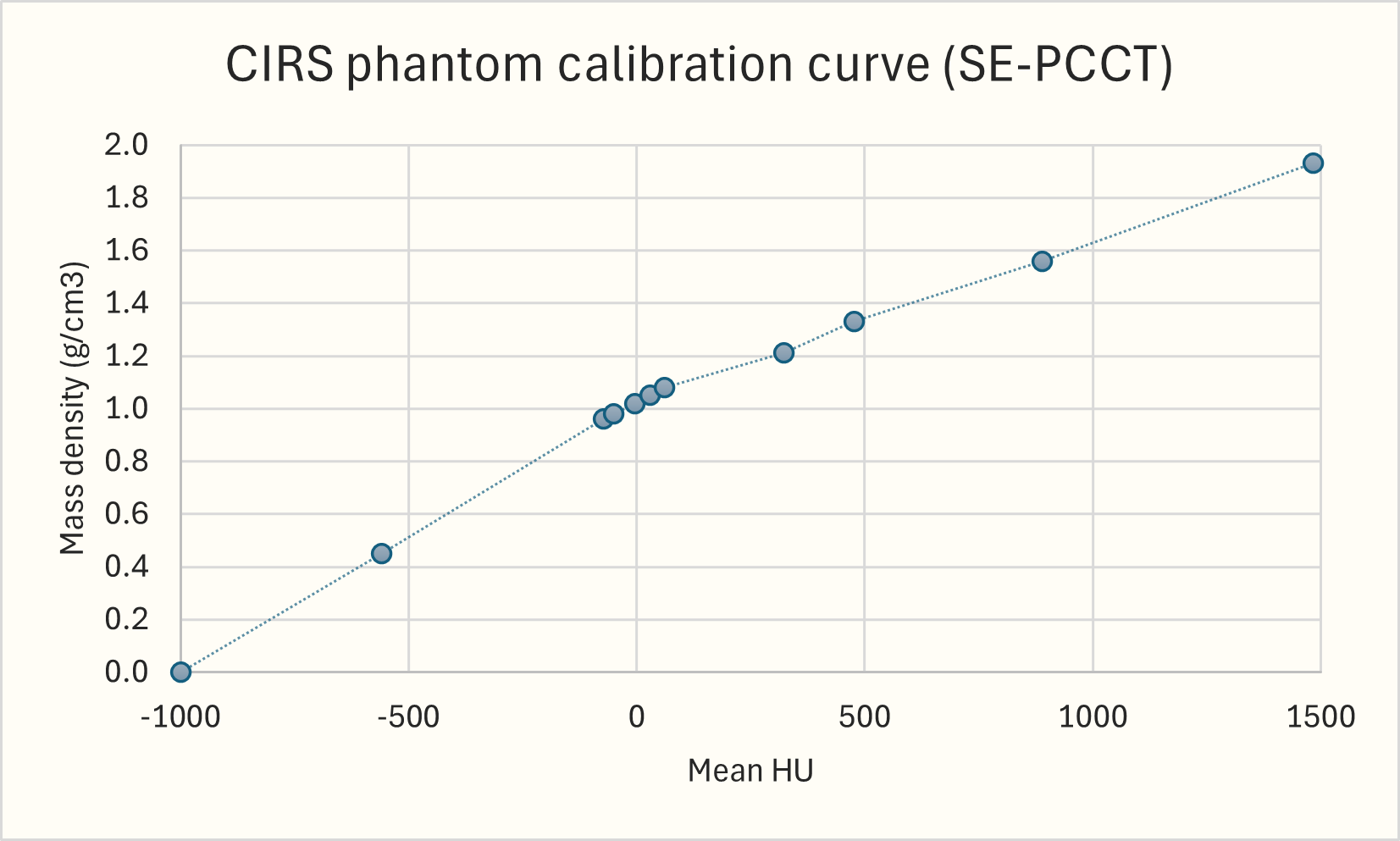}
    \caption{Linear HU-mass density calibration curve (120 kVp, 20 keV threshold) based on CIRS phantom inserts used for commissioning of the single energy photon counting CT.}
    \label{fig:calibration}
\end{figure}

Phantoms in figure \ref{fig:phantoms} were rescaled to match the size of CT reconstructions using the procedure described in section \ref{sec:gt_prep} and segmented regions were used to measure obtained SPR values from CT scans using the two different methods, the SE-PCCT plan and \textit{TissueXplorer} PCCT plan. The calibration curve used for the SE-PCCT plan was obtained from the tissue inserts arranged as in figure \ref{fig:gammex}. The shape of the curve is given in the figure \ref{fig:calibration}.

Table 2 contains ground-truth and estimated SPR values for different regions of interest in the XCAT head phantom. The SPR values in the \textit{TissueXplorer} PCCT plan were obtained directly from spectral data, while the ground-truth values were computed using equation \ref{eq:spr}. The "Outer body" region in table \ref{tab:spr_comparison} contains all tissues in the three-dimensional volume captured by the CT scan. The mean value of voxel-wise percentage difference in SPR across the "Outer body" is 0.28 \%.  In the critical structures considered as OARs, the smallest mean percentage difference was for the left optical nerve (0.29 \%) and the largest for the optic chiasm (6.60\%).

\begin{table}[ht!]
\centering
\caption{Comparison of ground-truth and TissueXplorer SPR values with standard deviation and percent error for selected structures.}
\resizebox{\textwidth}{!}{%
\begin{tabular}{lcccc}
\toprule
\textbf{Structure} & \textbf{Ground-truth SPR} & \textbf{TissueXplorer SPR} & \textbf{Standard deviation} & \textbf{Percent error} \\
\midrule
Eye (left)             & 1.049 & 1.037 & 0.027 & 1.14\% \\
Eye (right)            & 1.049 & 1.038 & 0.027 & 1.05\% \\
Optical nerve (left)   & 1.041 & 1.038 & 0.305 & 0.29\% \\
Optical nerve (right)  & 1.059 & 1.053 & 0.311 & 0.57\% \\
Optic chiasm           & 1.484 & 1.582 & 0.240 & 6.60\% \\
Nasal cavity           & 0.426 & 0.424 & 0.619 & 0.47\% \\
Brainstem              & 1.028 & 1.034 & 0.015 & 0.58\% \\
{\bf Outer body}       & 1.067 & 1.064 & 0.309 & 0.28\% \\
\bottomrule
\end{tabular}%
}
\label{tab:spr_comparison}
\end{table}

\begin{figure}[ht!]
  \centering
  \subfloat[\label{fig:gt_plan}]{
    \includegraphics[width=0.31\linewidth]{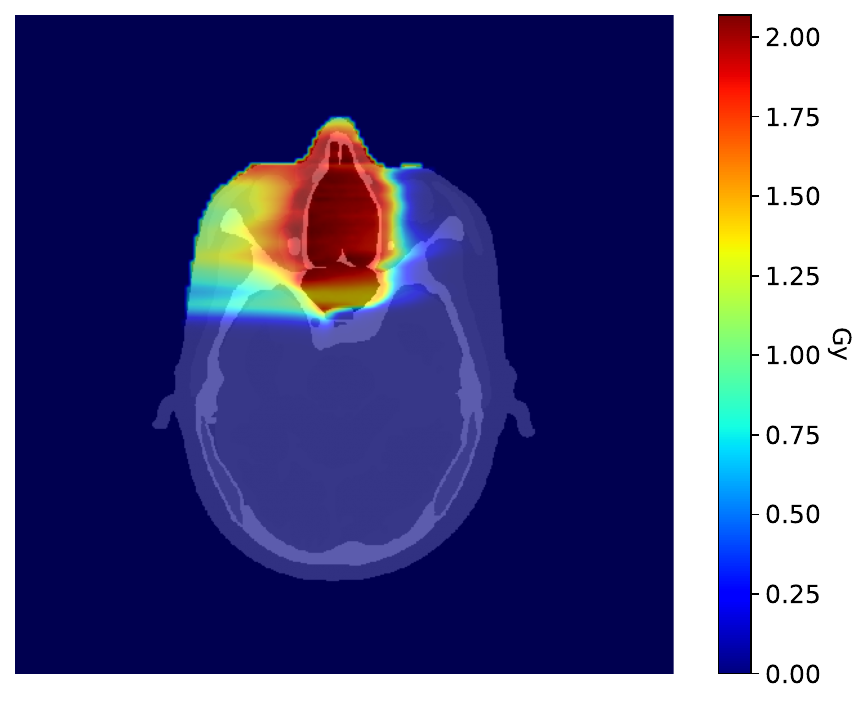}
  }\hfill
  \subfloat[\label{fig:HU_plan}]{
    \includegraphics[width=0.31\linewidth]{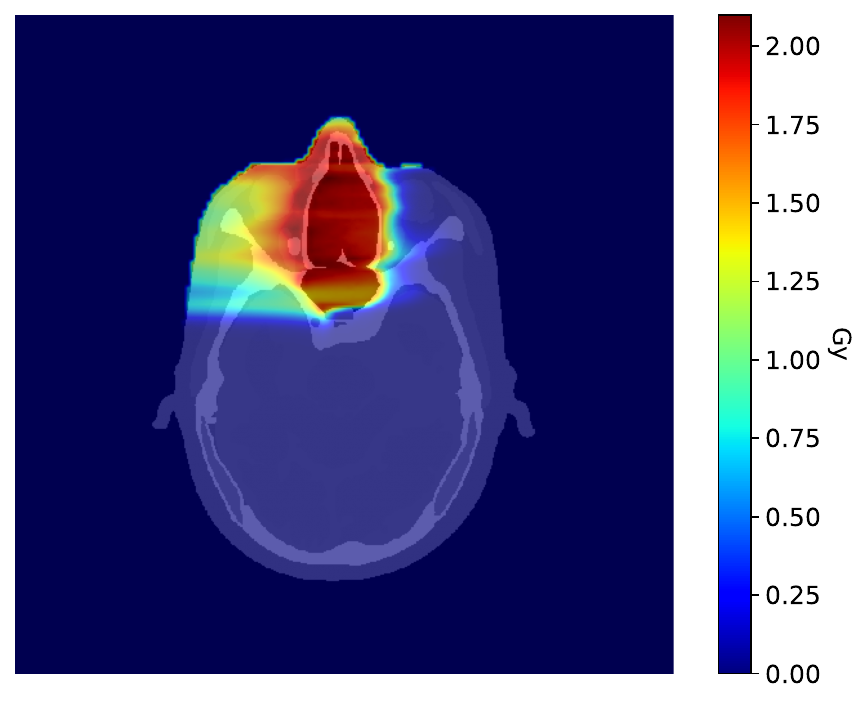}
  }\hfill
  \subfloat[\label{fig:TE_plan}]{
    \includegraphics[width=0.31\linewidth]{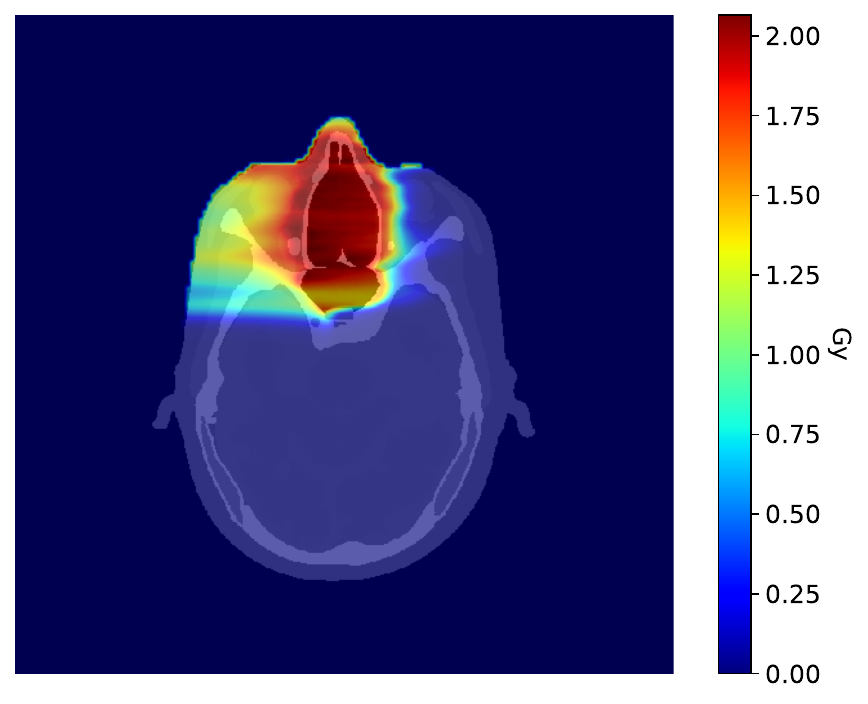}
  }\\[1ex]
  \subfloat[\label{fig:gt_bs}]{
    \includegraphics[width=0.31\linewidth]{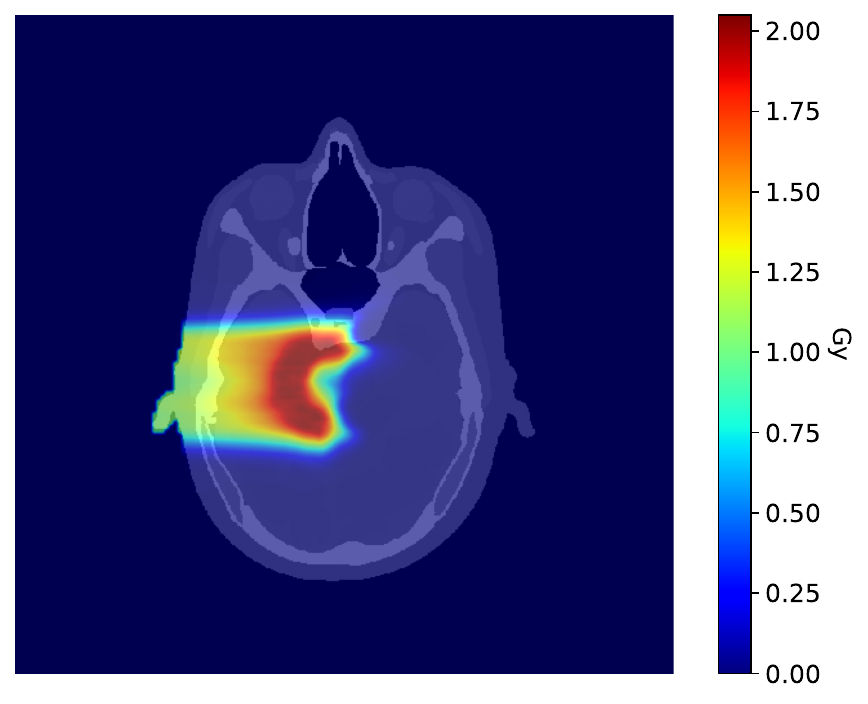}
  }\hfill
  \subfloat[\label{fig:HU_bs}]{
    \includegraphics[width=0.31\linewidth]{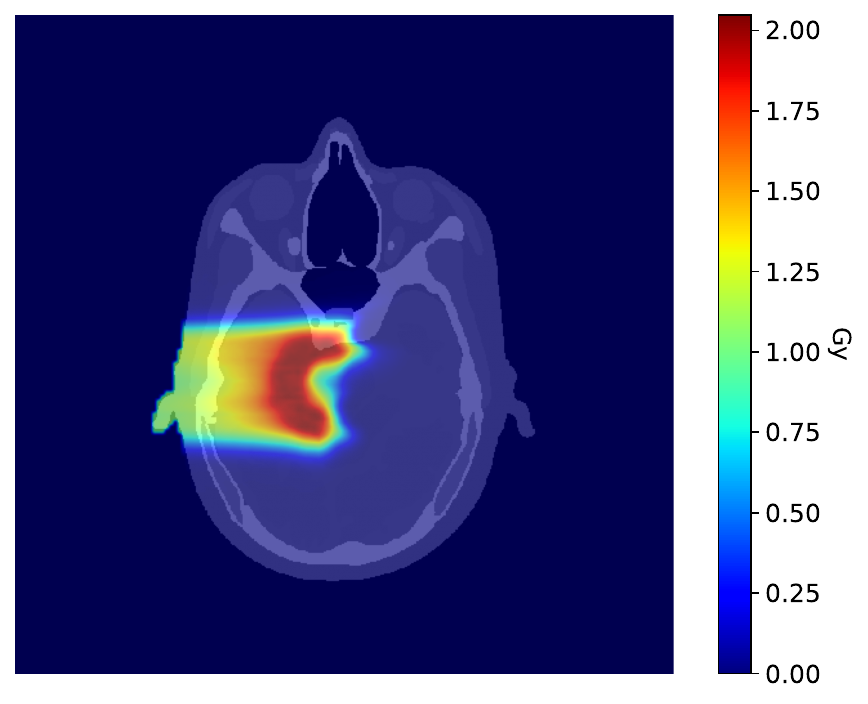}
  }\hfill
  \subfloat[\label{fig:TE_bs}]{
    \includegraphics[width=0.31\linewidth]{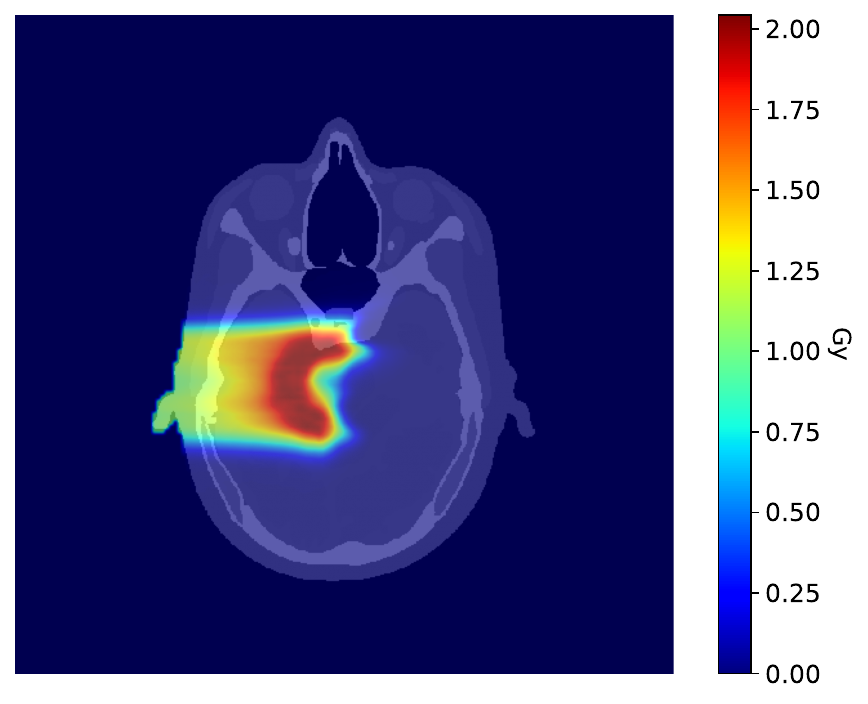}
  }
  \caption{Dose distribution in XCAT head phantom ground-truth plan obtained using known material compositions in table \ref{tab:composition} and equation \ref{eq:spr} for (a) nasal tumor CTV and (d) brain tumor CTV; recalculated SE-PCCT plan obtained using stochiometric calibration curve in figure \ref{fig:calibration} for (b) nasal tumor CTV and (e) brain tumor CTV; and recalculated PCCT plan obtained with \textit{TissueXplorer} software as described in section \ref{sec:tissueXplorer} for (c) nasal tumor CTV and (f) brain tumor CTV. The total dose is 2 Gy (RBE)  single-fraction, and the colorbar unit is in Gy.}
  \label{fig:treatment_plan}
\end{figure}

Figure \ref{fig:treatment_plan} gives the dose distribution for the 2 Gy (RBE) fraction in the ground-truth (figure \ref{fig:gt_plan} and \ref{fig:gt_bs}), and recalculated SE-PCCT plans using the stochiometric calibration curve (figures \ref{fig:HU_plan} and \ref{fig:HU_bs}) and \textit{TissueXplorer} software (figures \ref{fig:TE_plan} and \ref{fig:TE_bs}).

\begin{figure}[ht!]
  \centering
  \subfloat[\label{fig:gt_se_pcct}]{
    \includegraphics[width=0.45\linewidth]{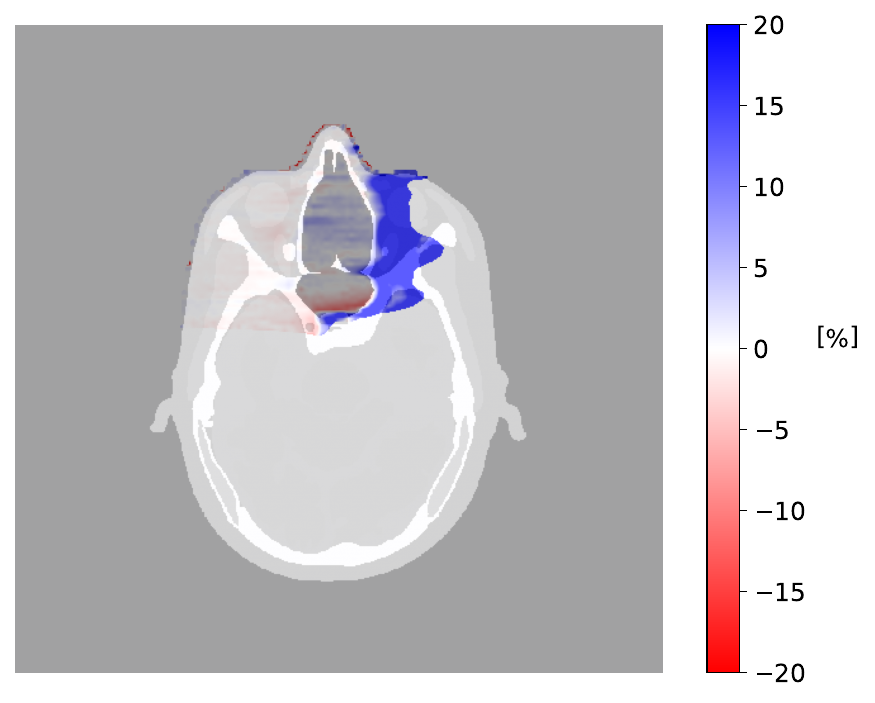}
  }\hfill
  \subfloat[\label{fig:tx_pcct}]{
    \includegraphics[width=0.45\linewidth]{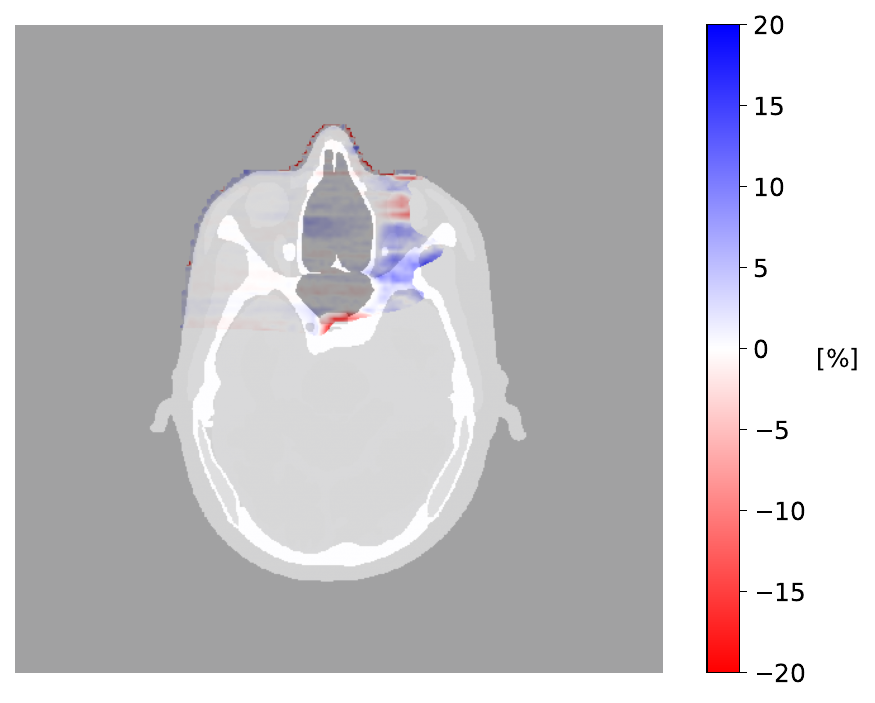}
  }\\[1ex]
  \subfloat[\label{fig:gt_se_pcct_bs}]{
    \includegraphics[width=0.45\linewidth]{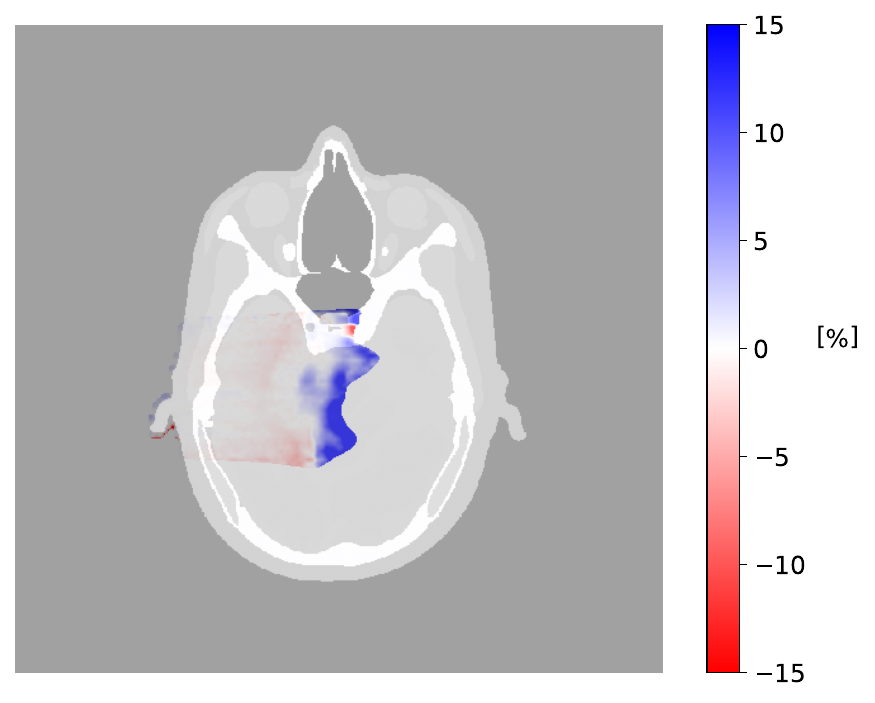}
  }\hfill
  \subfloat[\label{fig:gt_tx_pcct_bs}]{
    \includegraphics[width=0.45\linewidth]{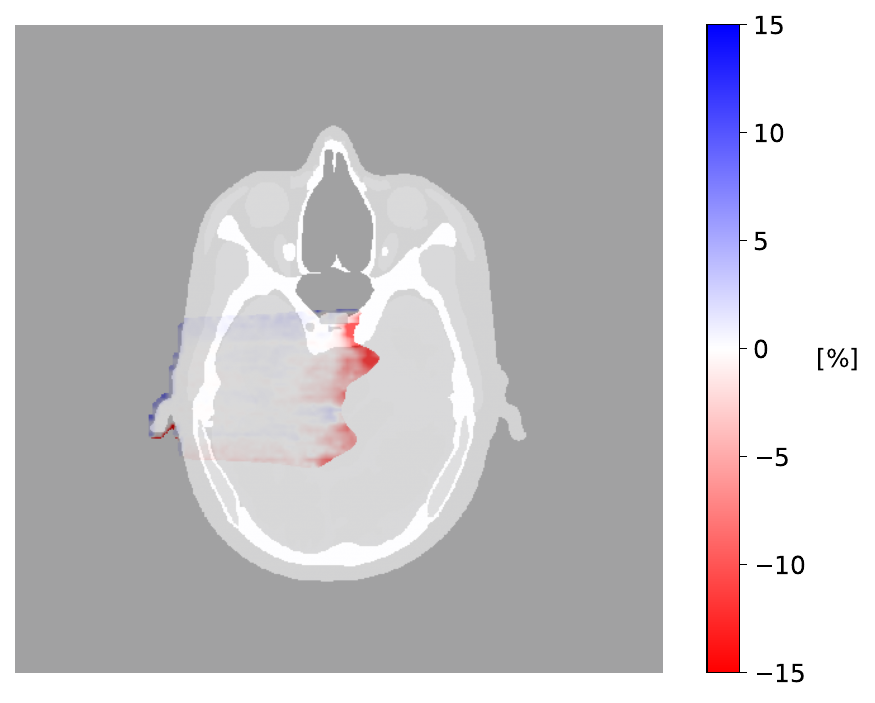}
  }
  \caption{Difference in dose distribution between ground-truth plan and recalculated single-energy photon-counting CT (SE-PCCT) plan for (a) nasal tumor CTV and (c) brain tumor CTV, and ground-truth plan and recalculated PCCT plan for (b) nasal tumor CTV and (d) brain tumor CTV. The scale bar shows percent differences.}
  \label{fig:plan_diff}
\end{figure}

To evaluate miscalculations in the dose distribution due to CT-to-SPR conversions, the voxel-wise percentage difference between recalculated plans and the ground truth plan is shown in the figure \ref{fig:plan_diff}. The blue color represents the dose undershoot, \emph{i.e.,} the regions with the dose lower than in the ground-truth plan, and red is the dose overshoot, \emph{i.e.,} the regions with the excessive dose compared to the ground-truth plan. The differences exist because of the miscalculation in the SPR values, and the largest differences are observed at the distal edge of the beam, arising from the miscalculations in the position of the Bragg peak. Qualitative comparison in the central slice shows that the SE-PCCT plan has significantly larger differences compared to the \textit{TissueXplorer} PCCT plan. 

\begin{figure}[ht!]
  \centering
  \subfloat[\label{fig:dvh_n}]{
    \includegraphics[width=0.75\linewidth]{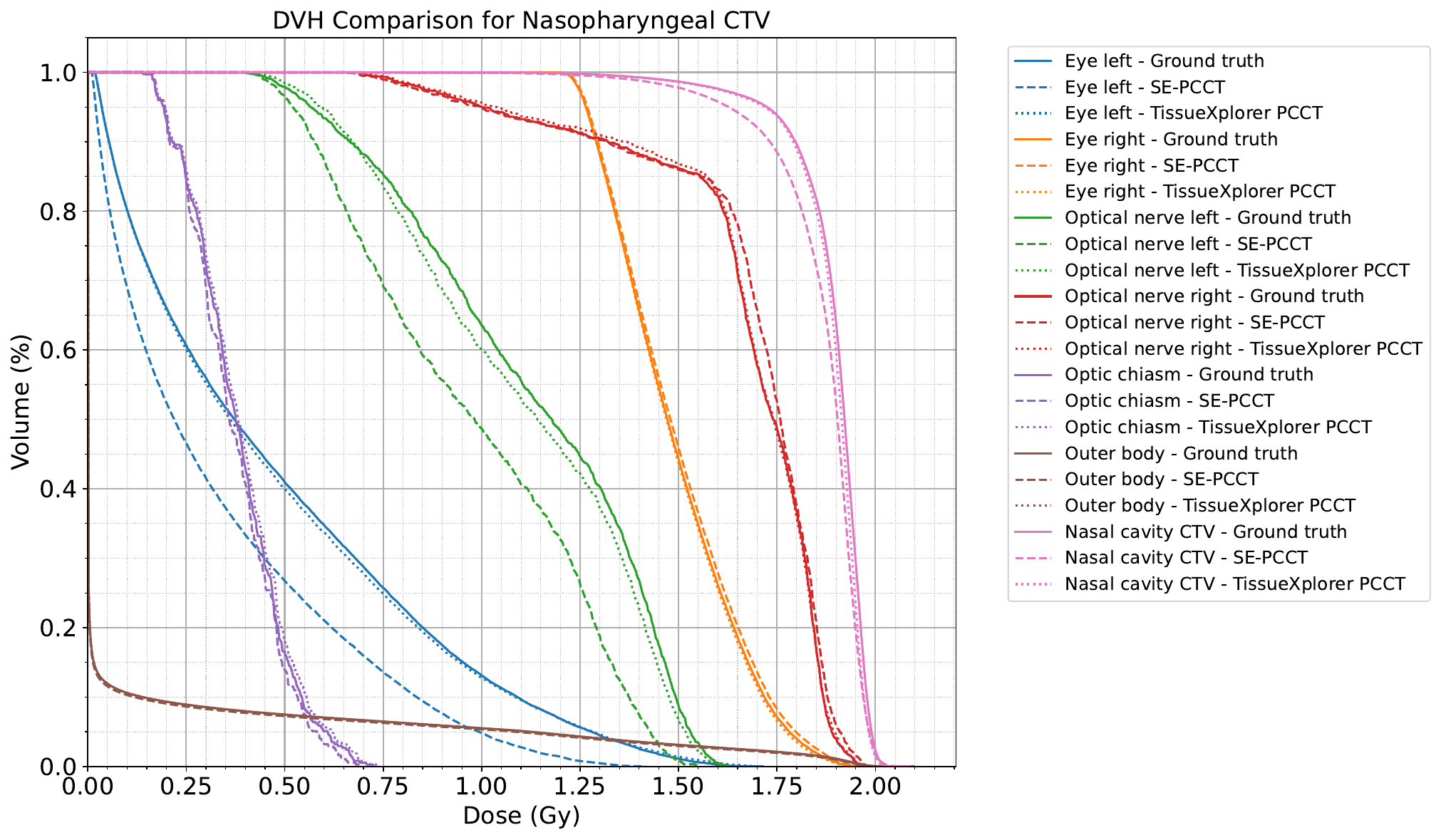}
  }\\[1.5ex] 
  \subfloat[\label{fig:dvh_bs}]{
    \includegraphics[width=0.75\linewidth]{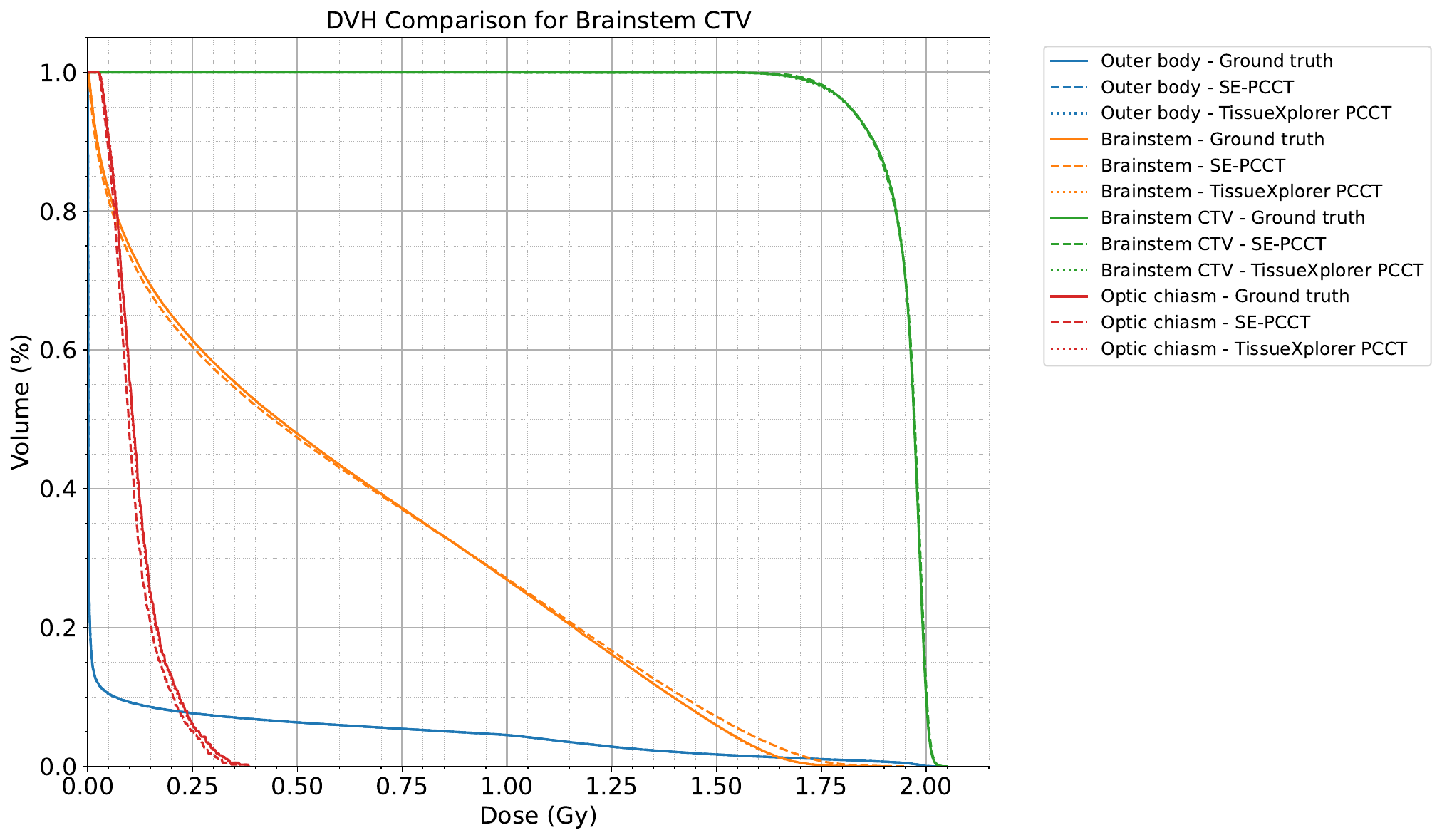}
  }
  \caption{Dose-volume histograms extracted from the ground-truth plan (full lines), single-energy photon-counting CT (SE-PCCT) plan with SPR obtained through the stoichiometric calibration curve in Figure~\ref{fig:calibration} (dashed lines), and the spectral PCCT plan with SPR obtained through the commercial prototype \textit{TissueXplorer} (dotted lines). Dose-volume curves for nasal tumor (a) are given for the left eye (blue), right eye (orange), left optical nerve (green), right optical nerve (red), optic chiasm (purple), outer body (brown), and the CTV in the nasal cavity (pink). Dose-volume curves for brain tumor (b) are given for the optic chiasm (red), outer body (blue), brainstem (orange), and the CTV in the proximity of the brainstem (green).}
  \label{fig:dvh}
\end{figure}

These results were confirmed by the dose-volume histograms comparison shown in figure \ref{fig:dvh}. The PCCT plan closely reproduced the ground-truth dose–volume distribution of both the CTV and OARs in both cases, with only minor deviations observed for the left optic nerve. In contrast, the SE-PCCT plan exhibited larger discrepancies across all delineated regions, most notably an underestimation in the CTV, left eye, and left optic nerve, and a dose overestimation in the right optic nerve. In the ground-truth plan, 99\% of the CTV received at least 1.43 Gy (RBE), the same as in the case of the PCCT plan, and 1.34 Gy (RBE) in the case of the SE-PCCT plan. The doses to the hottest 1 \% of the left eye were 1.49, 1.50, and 1.18 in ground-truth, PCCT, and SE-PCCT plans, respectively. When the target was the brain CTV rather than the nasal CTV, all three plans showed comparable dose distribution. Still, larger discrepancies from the ground truth plan were observed for the SECT plan than for the PCCT plan, as shown in figures \ref{fig:gt_se_pcct_bs} and \ref{fig:gt_tx_pcct_bs}.

\section{Discussion}

In this work, several critical aspects of range determination in proton therapy have been explored: the potential of virtual imaging as validation tools, the benefits of emerging PCCT technology, and the potential of increased accuracy of SPR prediction from spectral PCCT data. The methodology was described, and the results were based on a single representative case of a nasal and brain tumor in a virtual head phantom. 

Foremost, the study highlights the known issue with calculating SPR values from CT scans that remains one of the bottlenecks to overall accuracy in dose calculation \cite{moyers2020physical}. Our findings showed how the mismatch in the SPR calculation manifests itself in the mismatch in the dose distribution in realistic patient geometry. \textit{DukeSim} and similar tools designed for virtual imaging trials, enable efficient generation of a large number of clinically relevant cases on realistic CT simulators of scanners used in clinical practice \cite{abadi_dukesim_2019}. The XCAT program \cite{segars20104d} supports efficient generation of computational, anthropomorphic phantoms with lesion insertion tools and other realistic scenarios, such as breathing motion. This could enable population-based evaluation of mismatch in dose distribution due to SPR calibration and beyond, with clinical consequences such as dose to OARs or differences in normal tissue complication probability.

Beyond using the virtual imaging to replicate standard clinical practice and validate its accuracy in realistic patient geometry, this work shows how emerging photon-counting technology can support efforts in reducing range uncertainty. The overall uncertainty from CT imaging can be separated into the uncertainties arising from imaging noise and CT artifacts, and the uncertainties arising from the conversion of CT number into SPR. The two plans shown in figure \ref{fig:HU_plan} and \ref{fig:TE_plan} were obtained from the same CT acquisition, thus the same total photon statistics. The overall percent error in SPR calculation using \textit{TissueXplorer} software applied to the PCCT virtual monochromatic images was 0.28\%, which is lower than the results reported earlier on a dual-energy CT scanner \cite{sarkar_evaluation_2023,longarino_potential_2022}. In previous studies, PCCT scanners have demonstrated comparable or better quantitative performance than energy-integrating detectors, owing to lower imaging noise and a more uniform spectral response \cite{flohr2020photon}. While these features alone might not be significant to SPR prediction \cite{nasmark_influence_2023}, the PCCT scanners also provide inherent spectral separation. Clinical PCCT scanners showed comparable or better spectral imaging performance when compared to conventional DECT in spectral tasks such as virtual monochromatic imaging or estimating tissue density and effective atomic number \cite{vrbaski2023quantitative}. Almost exclusively, these quantities are obtained as a post-processing step of the material decomposition procedure, formally introduced by Alverez and Makovski \cite{alvarez_energy-selective_1976}. The effective atomic number has been successfully used to improve the stochiometric calibration curve \cite{yang_theoretical_2010}, however, effective atomic number is not a uniquely defined physical quantity. Multiple definitions and weighting schemes have been proposed depending on the energy range, interaction cross sections, and underlying assumptions \cite{vrbavski2023characterization, vrbaski_zcompare_2022}. As a result, calibration methods relying on the effective atomic number are often scanner- and vendor-specific. In contrast, virtual monochromatic images represent directly measurable, energy-dependent attenuation coefficients with inherently lower noise levels than effective atomic number maps. They provide a more physically grounded and vendor-independent basis for quantitative imaging and material characterization. It has been shown before that range uncertainty can be reduced using the virtual monochromatic images \cite{zhu_investigation_2023}.

\textit{TissueXplorer} can be regarded as an extension of the dictionary-based framework already implemented in commercial treatment planning systems, such as \textit{RayStation}. In current practice, these systems perform material selection based on CT numbers or derived mass density values, subsequently assigning each voxel to one of the predefined materials from an internal database \cite{janson_treatment_2024}. In contrast, the improved accuracy of the dose distribution in figures \ref{fig:plan_diff} and \ref{fig:dvh} arises from the incorporation of spectral information. By leveraging energy-dependent attenuation characteristics, \textit{TissueXplorer} enables a more precise material differentiation and quantification than the SE-PCCT calibration approach in the presented case, while remaining compatible with the existing treatment planning workflow. The improvements in SPR calculation have both clinical and economic relevance. More accurate SPR maps could allow reduced safety margins in treatment planning, potentially leading to improved dose conformity and better patient outcomes, as suggested in previous work \cite{taasti_clinical_2023}. 

Several aspects of this study need further investigation. Namely, the comparison presented showed differences in a single fraction for a single 2 Gy (RBE) lateral beam. In standard clinical cases, multiple fractions and different beam angles are used to cover the CTV and avoid damage to surrounding OARs as much as possible. Using multiple beam angles can reduce the dosimetric impact of uncertainties in the SPR, as the distal fall-off of individual Bragg peaks occurs at different spatial locations. Consequently, range errors that would otherwise cause localized dose degradation in single-field plans could be averaged out in multi-field configurations. Thus, comparison in a more realistic scenario will be the scope of our future studies. Additionally, some caution should be applied to the choice of tissue composition definitions in table \ref{tab:composition}. Although it has been taken from a well-established database for tissue composition, population-based differences in compositions for the same tissue type, as well as the differences in tumor tissue, might exist, which could have a nuanced effect on estimated SPR values with \textit{TissueXplorer}. The variation in tissue composition can be explored in population-based studies where variations are introduced during the phantom generation process. 

\section{Conclusion}

Virtual imaging framework offers an alternative approach to validate the accuracy of the treatment plan due to the HU-to-SPR conversion or other CT imaging-related errors in proton therapy. Compared to the conventional techniques, this approach offers a comparison of the differences in dose distribution for complex three-dimensional patient geometries. The vendor-agnostic software \textit{TissueXplorer} applied to photon-counting CT simulated head scan produced dose distributions that more closely matched the ground truth than those derived from the conventional stoichiometric calibration curve applied on the same scanner model for the test case of nasal and brain tumors.

\section*{Acknowledgments}
The authors thank Jovan Lovrenski, PhD, MD, and Olivera Vujinovic, PhD, from Vinaver Medical, for fruitful discussions and support. Authors acknowledge the HITRIplus (Heavy Ion Therapy Research Integration) project, funded by the European Union’s Horizon 2020 research and innovation program under grant agreement No. 101008548, for fostering collaboration and knowledge exchange that supported this work. Authors also acknowledge National Institutes of Health (P41EB028744 and R01EB001838).

\section*{Conflicts of interest}
Authors S. V. and G.S. have a relationship with Vinaver Medical. 
Unrelated to this study, author E.A. has a relationship with Siemens, GE, and Silomedics. E. S. has relationships with GE, Siemens, Imalogix, 12Sigma, SunNuclear, Metis Health Analytics, Silomedics, Cambridge University Press, and Wiley and Sons. Other authors declare that they have no known competing financial interests or personal relationships that could have appeared to influence the work reported in this paper.

\bibliographystyle{medphy.bst}
\bibliography{Proton_therapy}

\end{document}